\documentclass[a4paper]{article}
\usepackage[dvips]{graphicx}
\def\beq{\begin{equation}}
\def\eeq{\end{equation}}

\begin{document}
\title{It is the ambiguity. (But only three generations)}

\author{Alejandro Rivero \thanks {\tt rivero@wigner.unizar.es} \\
{\it Dep. Economia, Univ. Carlos III Madrid }}

\maketitle
\begin{abstract}
It is suggested that generations are linked to the need of calculating
curvature of space via a deformed or discrete calculus. Quantization would
limit the deformation, building three generations, and not four, as other
interpretation could imply. 
\end{abstract}

\section{Introduction}

It is known \cite{dowker,g} that the usual ambiguity 
in the definition of (partial) derivatives
of a function becomes a source of problems when we go to quantum theory.
The simplest example is Feynman quantum mechanics (or 0+1 dimensional
field theory), where different elections of discretisation for the
derivative drive to different ordering rules in the quantized theory; Weyl
corresponds to the symmetric one, Born-Jordan to the forward derivative, and
so on, and even more exotic effects could be got by gauging the ambiguity.

My paper \cite{generations} has been understood in \cite{four} as if it
were giving
theoretical support to the existence of a fourth generation. Indeed, as
it is pointed in \cite{generations} and also previously in \cite{conjectures},
I believe that the ambiguities of a discrete curvature only point to
three generations, the extant ambiguity being absorbed in a scale parameter
which relates to Plank constant (and, to the deformation parameter of the
calculus). It is my fault that this process is just half cooked, in 
sparse references at the end of \cite{fletter} and, as appendix, in
\cite{grupoid}. I apologize I can not help with a detailed example yet, so in
order to try to clear the confusion, this paper can only to expand on
the ideas of our previous \cite{generations}. 

So, please consider this note
as a traditional conference poster, trying to put illustrations
to the previous work. 
The examples of the previous papers were formulated in the context
of non commutative geometry. Here we will keep ourselves in the  
conceptual cloud. 

\section{Pictorial image}

\begin{figure}[!htb]
\begin{center}
\includegraphics{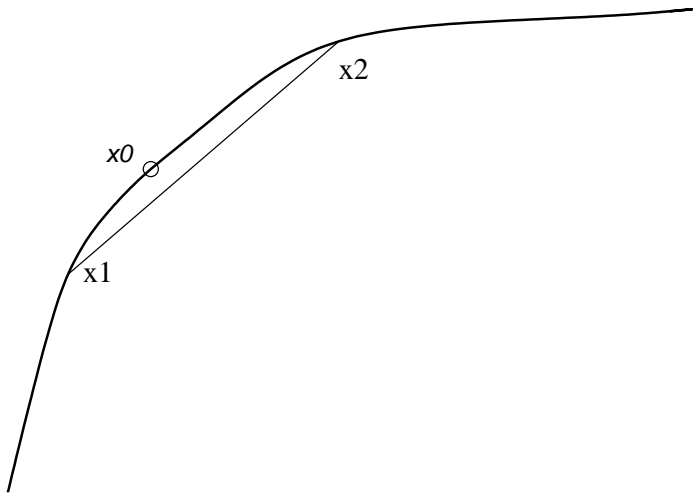}
\caption{q-tangent vector to the point $x_0$}
\label{tangente}
\end{center}
\end{figure}

The main observation, figure (\ref{tangente}), is that to define
a vector tangent at a point $x_0$ we need to build two series
of points approaching $x_0$, so the limit $(x_2-x_1)/\epsilon$ will
give the tangent vector. It seems that there are two ambiguities,
to choose $x_2$ and to choose $x_1$, but one of them can be
absorbed into the scale parameter. The extant ambiguity is the one
we proposed to consider a mass. 

\begin{figure}[htb!]
\begin{center}
\includegraphics{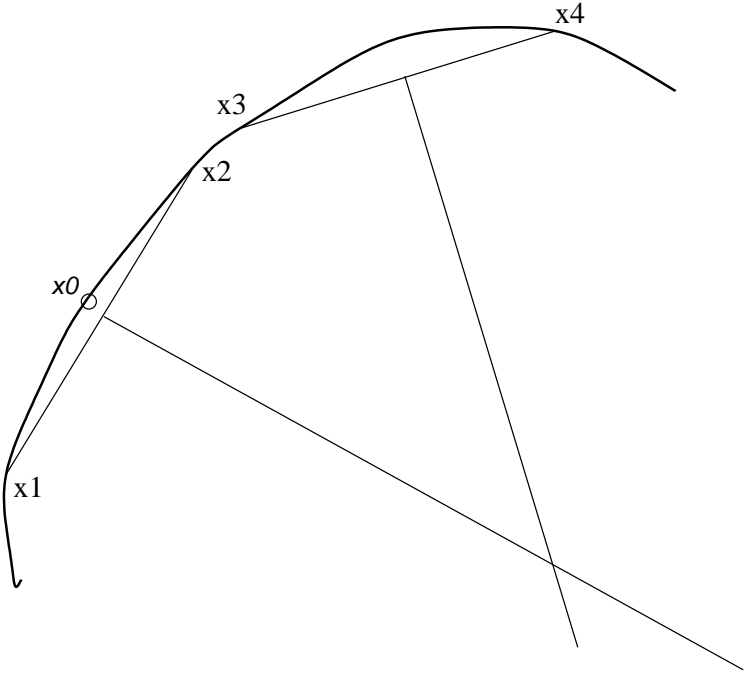}
\caption{Curvature of a curve through $x_0$}
\label{radio1d}
\end{center}
\end{figure}

Consider now, as in figure (\ref{radio1d}), a curve of which we want to
know the curvature at a point. This implies to give four points, so
the orthogonal to two q-tangent vectors will cross marking the position
of the radius of curvature, and giving us the inverse of the curvature
when the continuous limit is approached.

Again the scale can absorb one of the ambiguities, and we have three
free parameters, that we can identify with three masses.

\begin{figure}[htb!]
\begin{center}
\includegraphics{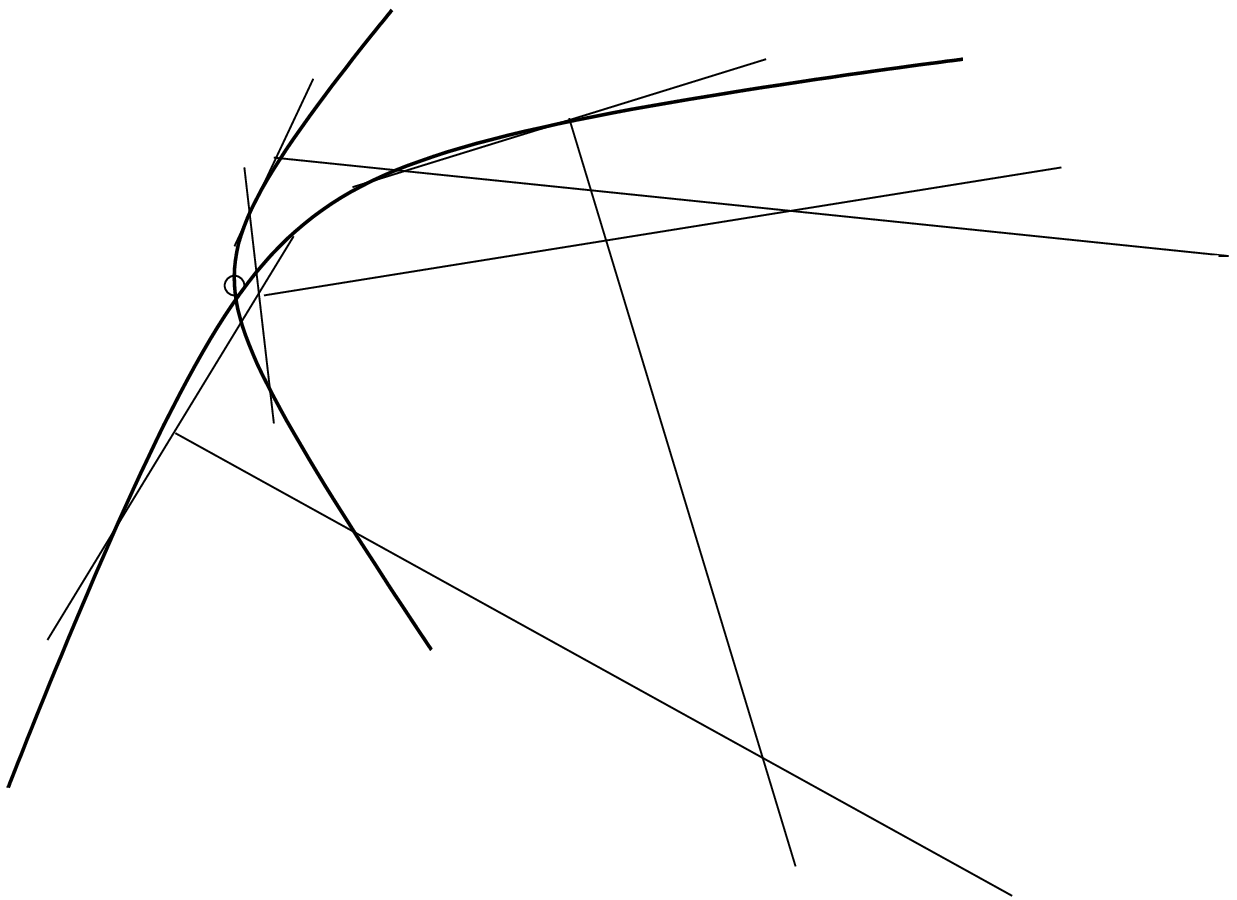}
\caption{Two curvatures of a surface}
\label{radio2d}
\end{center}
\end{figure}

When going to higher dimensions, the play is not more ambiguous, but
it is more complicated (Figure \ref{radio2d}). The curvature tensor
is build from the curvatures of the geodesic surface tangent
to each 2-plane. Every direction must be considered, and dependences
between the ambiguities of orthogonal directions are not clear.

There are ways to simplify the task, for instance asking for
additional restrictions (isotropy, homogeneity) to the space-time
manifold. In any case, we should expect now to multiply our
degeneration times the number of dimension of space time, to be
able to cope with every curvature. Thus we will get four particles
(from space-time dimension) and three generations (from ambiguity).  

\begin{figure}
\begin{center}
\includegraphics[angle=15]{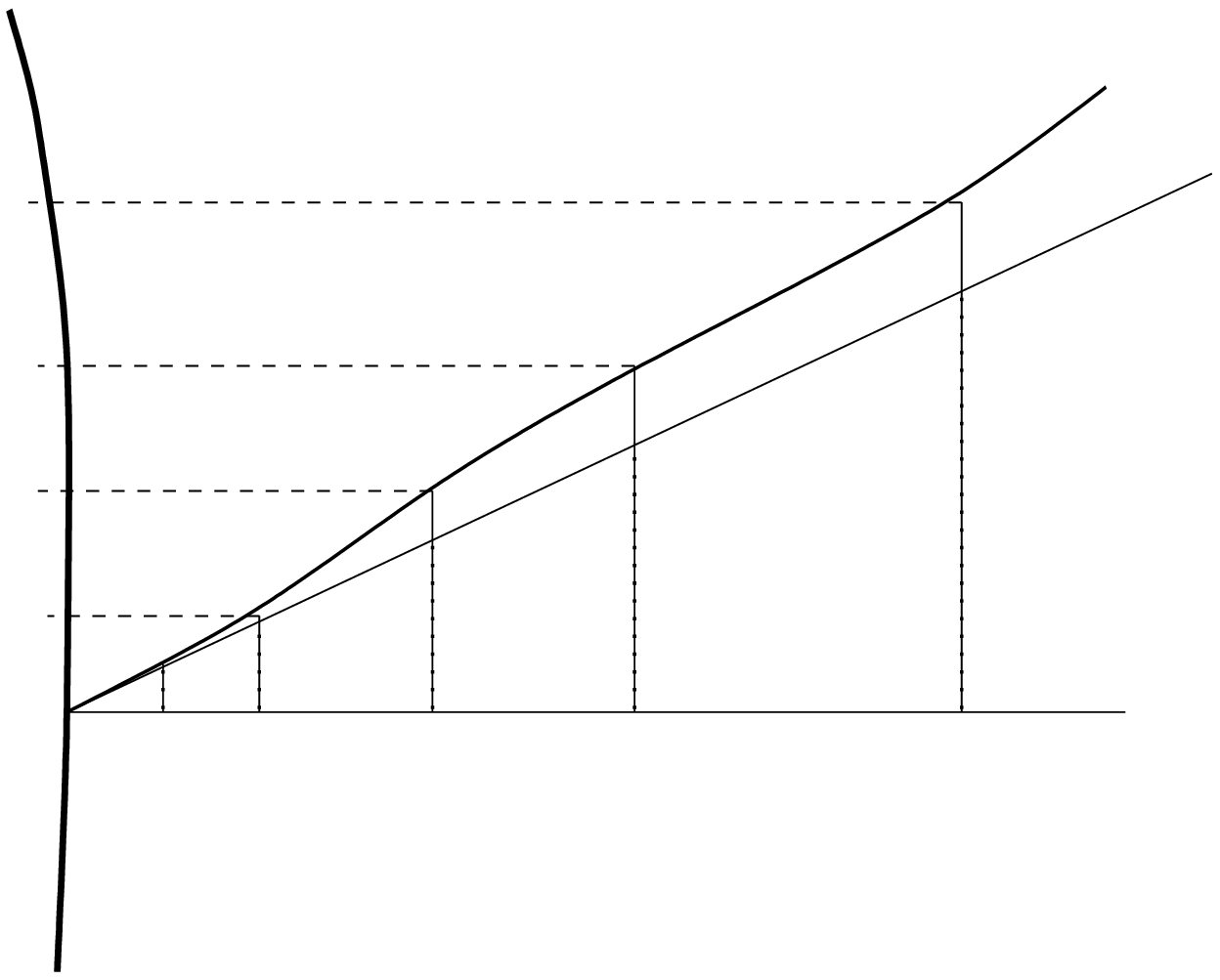}
\caption{Tangent vector as a limit or "instantaneous velocity"}
\label{tales}
\end{center}
\end{figure}

\begin{figure}
\begin{center}
\includegraphics{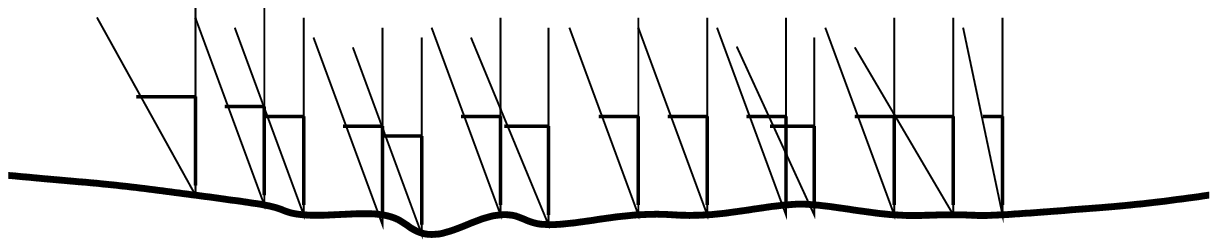}
\caption{Summation (integration) across a section of the tangent
bundle. Should a length scale be defined at every point?}
\label{suma}
\end{center}
\end{figure}

Last, consider how the scale parameter could come into play. Every
tangent vector is defined via tales theorem, figure (\ref{tales}), relating
the time (or parameter of the curve) with the distance. Classically
each triangle is well defined at each point of the curve. Now, doubts
can be raised when we make an integration (figure \ref{suma}) in the
"deformed" way, before any limit. It seems that we should introduce
an scale parameter $\epsilon$, to be able to add the quantities of each
triangle, and that the continuous limit should come when $\epsilon \to 0$.

The process as I see it is a little more involved, as first one needs
to use scale invariance to go from a bare $\epsilon$ to a 
renormalized $h$, and the classical limit is $h\to 0$. This method
is needed because the groupoid of paths given by Connes does not
add $(x,y,t)(y,z,t)$ to $(x,z,2t)$ but to $(x,z,t)$.  

\section{Interdisciplinary work}

The most popular of all the regularisation problems is the one of 
fermion doubling, which
happens when we try to quantize fermions in lattice quantum field theory.
Here, if we use only the symmetric definition of derivative we finish with 
a set of $2^d$ fermions for each initial fermion in the theory. Wilson cures
this by incorporating the ambiguity to the theory, building each derivative
as a combination of symmetric and antisymmetric part, and then giving a
high mass to the antisymmetric combination, which controls the unwanted
degrees of freedom. Recently \cite{luscher, fujikawa} Wilson' approach has
been resuscited and the Ginsparg-Wilson condition is looking for
a place in non commutative geometry. L\"uscher' approach is perhaps closest to
Dimakis or Majid ones, but Balachandran \cite{fuzzy} is already looking 
for a role for it
near to the axiomatic of NCG manifolds.

From the point of view of non commutative geometry, it seems 
\cite{almostnogo} that naive lattice
field theory does not qualify as a differential geometry, as even when
multiple copies of elements are taken (which seems needed to cope with
NCG first order axiom), it fails to fulfill Poincare duality. To get out
of this trap, the suggestion should be to act with the Dirac
operator in different points of the space.
 
We will do this by introducing small finite differences
between the fields attached to each fermion, and relating this difference
to the inverse of the mass reasoned in the pictorial show,
 so that in the very low energy limit the
implemented difference becomes equal to the derivative it discretizes.
Our goal 
is to expand NCG Lagrangian to contain information about
quantization ambiguity, but with this technique we also get to introduce the
spectrum of masses. 

Time ago in Barcelona, Alain, in a dual session with Asthekar, suggested that
the newer versions \cite{reality, gravity} of the Connes-Lot approach 
should be seen as an low energy approximation to a completely
non commutative space only visible at high energy. In some sense, here
the methodology is reversed, betting first of a "very non
commutative" model and trying to guess a method to go down to
low energy. At very low energy, only gravity should be seen, while
at intermediate energy, Connes-Chamseddine or Connes-Lot should
be suitable approximations. 

Just to visualize such approximation we want to keep ourselves
using the Dirac operator formalism. On the other hand,
from the Tangent Groupoid construction, we know that the set of functions
over the tangent space $TM$ of a manifold can be pasted, via Weyl
quantization \cite{pepin, connes}, to the set of kernels
$k(x,y)$ of operators in a Hilbert space, and this formalism
is very near to the finite difference scheme proposed above. 
It should be nice if
 our candidates for differential
forms had some duality with this space of operators. Also because
the tangent groupoid seems closer to q-deformations as made
by Majid and others, and to the non-commutative formalism
used by lattice theoretists in the above referred works.

\section{Mass}

Basic mass, as we have seen, should be fixed by the position of the vector
of figure \ref{tangente} respective of the point which "differentiation" is
assigned to.

For each particle, a mass relationship can be imposed asking to the second
derivatives to give the same result as a iteration of the first derivatives. So
the two vectors of a curvature would be given by the positions of the extreme
points of the vector giving the first  derivative.

Such relationship could be not needed if some geometrical consistency conditions
where imposed to the mass matrix. For instance, it is know that Poincare 
Duality forces the Connes-Lott model generations to be degenerated in mass.
More restringent conditions could appear.

Finally, the mass relation between different particles is the touchiest point. 
My bet is to link it to a preferred kind of metrics, with a preferred set
of coordinate systems. Very much as it happens in a spherical set of
coordinates: a variation across $\delta r$ has no additional weight, variations
across  $\delta\theta, \delta\phi$ carry an additional weight $r$, and variation across $\delta\phi$ carries an additional term respect to $\delta \theta$. 
In \cite{conjectures}, this was stated with an obscure comparison between
quarks and angles.

\section{Acknowledgements}

To gauge the ambiguity is a suggestion of E. Follana, yet to explore.
 Cheerfully acknowledged, as well as a lot of support to discuss other
ideas. Momentum space was seen as configuration space
with a field of forces in some coffee talks with J. I. Martinez, 
according my old notebooks. Again, yet to explore. And, as noted elsewhere,
 the main thesis of
this paper surfaced while a bedroom talk with J. Guerrero at Vietri, where
I was driven to think in analogies between the four fundamental fermions
and a four dimensional volume form. More about this can be forthcoming
in \cite{junk}, where we wonder about the relation between junk removal
and the Pauli antisymmetrization of $N$ fermions. 

The relaxing ambiance provided by the folks of {\it La Latina}, the Madrid 
inner neighborhood where I am doing a half sabbatical,
should also be acknowledged.


\begin{thebibliography}{50}

\bibitem{fuzzy} A. P. Balachandran et al., {\it The Fermion Doubling Problem
and Noncommutative Geometry}, hep-th/9911087
\bibitem{four} Chao-Shang Huang et al., {\it The $B\to X_sl^+l^-$ and $B\to X_s \gamma$ decays with the fourth generation}, hep-ph/9911203
\bibitem{pepin} JF Cari\~nena et al., {\it Connes' Tangent Groupoid and
Deformation Quantization}, J. of Geom. and Phys, v 32 (1999). math/9802102
\bibitem{connes} A. Connes, {\it Non Commutative Geometry}, Academic Press 1994
\bibitem{reality} A. Connes, {\it Non Commutative Geometry and Reality}
\bibitem{gravity} A. Chamseddine, A. Connes, {\it A Universal Action
Formula}, preprint hep-th/9606056 
\bibitem{axioms} A. Connes, {\it Gravity coupled with matter and the
foundation of non commutative geometry}, preprint hep-th/9603053
\bibitem{dowker} J.S. Dowker,  Path Integrals and Ordering
Rules, {\it J. Math. Phys.} {\bf 17} (1976)
\bibitem{fujikawa}, T. Fujikawa et al., {\it Non-commutative Differential
Calculus and the Axial Anomaly in Abelian Lattice Gauge Theories},
hep-lat/9906015
\bibitem{almostnogo} M. G\"okeler and T. Sh\"uker, {\it Does 
noncommutative geometry encompass lattice gauge theory}, hep-th/9805077
\bibitem{g} Roy R. Gould, {\it Am. J. Phys.}, {\bf 63}, n. 2 (1995)
\bibitem{luscher} M. L\"uscher, {\it Chiral gauge theories on the lattice
with exact gauge invariance}, hep-lat/9909150
\bibitem{majid} S. Majid,{\it Advances in Quantum and Braided Geometry}, 
 q-alg/9610003 v2
\bibitem{grupoid} A. Rivero, {\it Introduction to the tangent 
grupoid}, dg-ga/9710026. 
\bibitem{fletter} A. Rivero, {\it A short derivation of Feynman
formula}, quant-ph/9803035
\bibitem{conjectures} A. Rivero, {\it Some conjectures looking of a NCG theory},
 hep-th/9804169
\bibitem{generations} A. Rivero, On generations, hep-th/9905021
\bibitem{junk} A. Rivero, {\it Junk: the fermionic ansatz}, work in course.
\bibitem{mariano} M. Santander, {\it Interpretacion geometrica de la gravitacion}, DFTUZ/93/11

\end{thebibliography}
\end{document}